\documentclass[twocolumn]{jpsj2}

\def\runtitle{Hole Dynamics in the Orthogonal-Dimer Spin System
}
\def\runauthor{Yasuhiro \textsc{Saito}, 
Akihisa \textsc{Koga} and Norio \textsc{Kawakami}
}

\title{
Hole Dynamics in the Orthogonal-Dimer Spin System
}

\author{%
Yasuhiro Saito, Akihisa Koga and 
Norio Kawakami
}

\inst{
Department of Applied Physics, Osaka University, Suita, 
Osaka 565-0871, Japan
}

\recdate{\today}

\abst{
The dynamics of a doped hole in the orthogonal-dimer spin system 
is investigated systematically in one, two and three dimensions.
By combining the bond-operator method with the self-consistent
 Born approximation, we argue that a dispersive quasi-particle
 state in the dimer phase is well defined 
even for quasi-two-dimensional systems. On the other hand, 
 a doped hole in the plaquette-singlet phase  hardly 
itinerates, forming an almost localized mode. We further
 clarify that although the quasi-particle weight in the dimer phase is 
decreased in the presence of the
interchain coupling, it is not suppressed but
even enhanced upon the introduction of the interlayer coupling.
}

\kword{%
$\rm SrCu_{2}(BO_{3})_{2}$, Orthogonal-dimer, Shastry-Sutherland model, self-consistent Born approximation
, bond operator}

\begin{document}
\sloppy
\maketitle

\section{Introduction}
Recently, low-dimensional quantum spin systems with frustration have 
attracted much interest. 
Among them, a quasi-two-dimensional (2D) spin gap compound
$\rm SrCu_{2}(BO_{3})_{2}$ \cite{Kageyama,Kodama} 
is outstanding in its unique features. 
The crystal structure of the compound is such that  
$\rm Cu^{2+}$ ions with spin $S=\frac{1}{2}$ are 
located on 
the orthogonal-dimer lattice,
resulting in the characteristic magnetic properties.
\cite{Onizuka,Aso,Nojiri,Lemmens,Kageyama2}
Miyahara and Ueda\cite{Miyahara} pointed out that 
the orthogonal-dimer lattice is equivalent to the square lattice 
with some diagonal bonds, which is 
referred to as the Shastry-Sutherland model.\cite{Shastry}
(See Fig. \ref{fig:model})
The system has been providing a variety of interesting
issues 
\cite{Miyahara-4},
e.g. quantum phase transitions,
\cite{Miyahara,Koga-1,Knetter,Weihong,Muller,Mila,Chung,Carpentier,Takushima,Sigrist2}
unique triplet excitations,\cite{TM,Momoi,Knetter,Totsuka,Fukumoto-2,Knetter-2,Munehisa} 
magnetization plateaus.
\cite{Momoi,Totsuka,Miyahara3,TM,Muller,Misguich,Fukumoto-pla,Miyahara7} 
It was found that the ground state of 
the compound $\rm SrCu_{2}(BO_{3})_{2}$
is in the dimer phase close to the quantum phase
transition point.\cite{Miyahara} 
Therefore, the application of the magnetic field
\cite{Kageyama,Onizuka} or the pressure\cite{Kageyama2}, could trigger 
a quantum phase transition. Also, some reports claimed that the 
compound $\rm SrCu_{2}(BO_{3})_{2}$ 
has a rather large interlayer interaction to explain the
 experimental data such as the susceptibility and the specific heat.
\cite{Miyahara2,Knetter}

The study of a single hole doped into the above orthogonal-dimer 
system is important,\cite{Vojta}
 since its dynamics determines the profile of
the photoemission spectrum.  In comparison with the extensive theoretical
investigations of static quantities, such one-particle spectral
properties have not been studied systematically, except for Vojta's work 
on the dimer phase of the 2D Shastry-Sutherland model.\cite{Vojta} 
In particular, it is desirable to clarify how the interlayer coupling,
which turns out to be important for the static quantities, 
affects the photoemission spectrum.
Moreover, such investigations of a doped hole may provide
a starting point to study the system with finite density of holes,
which may be realized by the chemical substitution of some
elements.  Although simple substitution of 
 $\rm Zn$ ions for  $\rm Cu$ ions results in
 localized magnetic moments,\cite{Dope} it is 
an interesting issue to realize a metallic orthogonal-dimer 
system experimentally.

Motivated by the above topics, we investigate 
the dynamics of a doped hole 
in the orthogonal-dimer system systematically in one,
two and three dimensions.
By means of the self-consistent 
Born approximation (SCBA)
\cite{rink,horsh,liu,frank,jurecka,Saito} 
with the bond-operator representation
\cite{jurecka,Saito,Sachdev,Sigrist,Vojta,Matsumoto} and 
also the exact diagonalization (ED), 
we discuss how the characteristic orthogonal-dimer structure affects 
the motion of a doped hole. 

This paper is organized as follows. In Sec. \ref{sec2}, we introduce 
the orthogonal-dimer spin system, which is relevant 
to describe the system of $\rm SrCu_{2}(BO_{3})_{2}$.
In Sec. \ref{sec3}, we briefly summarize the 
 bond operator representation for quantum spins, which is combined with 
the self-consistent Born approximation to treat a doped hole.
This method is used for the analysis of the dimer phase, whereas 
the ED method is also employed to treat the hole
dynamics in  the plaquette phase.
In Sec. \ref{sec4}, we study the dynamics of a doped hole 
systematically in  one, two and three dimensional orthogonal-dimer
systems, and clarify the role of the lattice geometry 
on the dynamics of a doped hole.  
A brief summary is given in Sec. \ref{Summary}.

\section{model}\label{sec2}
\begin{figure}[htb]
\begin{center}
\includegraphics[width=7cm]{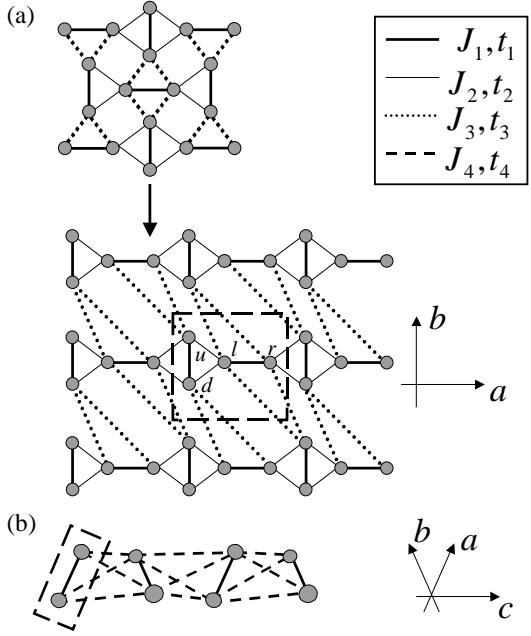}
\end{center}       
\caption{Orthogonal-dimer 
system for $\rm SrCu_2(BO_3)_2$, which has alternating 
orthogonal-dimers
in all directions along the $a$, $b$  and $c$ axes:
(a) the 2D  Shastry-Sutherland model in the $a$-$b$ plane.
Square unit cell containing four spins 
is shown as a long dashed line.  If we set $J_3=t_3=0$
(dotted line), the system is reduced to an
orthogonal-dimer chain with alternating dimer-plaquette 
structure; (b) configuration of the dimers along the $c$-axis, 
which forms another type of orthogonal-dimer chain.
This is equivalent to a ladder model with additional
exchange interactions for diagonal bonds, which is referred to as 
tetrahedra chain.
}
\label{fig:model}
\end{figure}

We investigate the orthogonal-dimer spin
model,\cite{Knetter,Ueda,Koga-2} which is 
schematically drawn in Fig. \ref{fig:model}. 
To deal with the system with a  hole,
we consider the following {\it t-J} Hamiltonian,
\begin{eqnarray}
H&=&H_{t}+H_{J},
\label{eq:hamiltonian}
\\
H_{J}&=&\sum_{\kappa=1,4}J_{\kappa}\sum_{<i,j>\in D_\kappa}{\bf S}_{i}\cdot
{\bf S}_{j}
\label{eq:hj}
\\
H_{t}&=&-\sum_{\kappa=1,4}t_{\kappa}\sum_{<i,j>\in D_\kappa,\sigma}
\hat{c}_{i,\sigma}^{\dag}\hat{c}_
{j,\sigma}
+H.c. ,\label{eq:ht}
\end{eqnarray}
with $\hat{c}_{i,\sigma}^{(\dag)}=[c_{i,\sigma}(1-n_{i,
-\sigma})]^{(\dag)}$, where $c_{i,\sigma}^{(\dag)}$ 
annihilates (creates) an electron and $n_{i,\sigma}$ is the electron 
number 
with spin $\sigma(=\uparrow, \downarrow)$ at the $i$th site.
The corresponding spin operator is defined by 
${\bf S}_{i}=\frac{1}{2}\sum_{\sigma,\sigma'}c_{i,\sigma}^\dag 
{\bf \tau}_{\sigma,\sigma'}c_{i,\sigma'}$, 
where ${\bf \tau}$ is the Pauli matrix.   We 
introduce four types of hopping integrals
$t_{\kappa}>0$ and exchange couplings  $J_{\kappa}>0 (\kappa=1,2,3,4)$,
which are specified by the bond pairs $D_\kappa=\{J_\kappa, t_\kappa \}$ 
shown in Fig. \ref{fig:model}. Note that
this generalized model has the orthogonal-dimer structure in 
all three directions,  thereby allowing us to
discuss the hole doping effect systematically in one, two
and three dimensions.
When $J_2=J_3$ and $J_4=0$, the system is reduced to 
the well-known 2D Shastry-Sutherland model,\cite{Shastry} 
static properties of which have been intensively studied so far.
\cite{Miyahara,Koga-1,Knetter,Weihong,Muller,Mila,Chung,Carpentier,Takushima,Sigrist2,
TM,Momoi,Totsuka,Fukumoto-2,Knetter-2,Miyahara3,Misguich,Fukumoto-pla}
The system is further reduced to a 1D 
orthogonal-dimer system if we consider the chain system
along the  $a$ or $b$ axis (lower panel of Fig. \ref{fig:model}(a)). 
Also, the 1D system along the 
 $c$ axis forms another orthogonal-dimer 
chain, as shown in Fig. \ref{fig:model}(b).
Magnetic properties of 
these chain models have been studied 
in detail. \cite{Ivanov,Schulenburg,Richter,Koga-3,Honecker,
Gelfand,Honecker,Brenig,Kim,
Totsuka-lad,Sutherland,Kawaguchi}

In contrast to well-studied static properties, systematic investigation of
one-particle spectral properties\cite{Vojta} is still lacking, which we 
will address in the 
following sections.

\section{Bond operator approach}\label{sec3}
In this section, we introduce a method 
to treat  the dynamics of a doped hole.
Since we are mainly concerned with the dimer phase, it is convenient to
 use the bond operator representation,
\cite{jurecka,Saito,Sachdev,Sigrist,Vojta,Matsumoto}  which may describe
the ground state as well as low-energy excitations in the 
dimer phase rather well.
The essence of this method is to represent the
Hamiltonian in terms of "bond operators", which act on each
dimer pair formed by $J_1$, instead of original operators 
for spins and electrons.
Since the generalized Shastry-Sutherland model has four sites in a 
unit cell, we introduce the bond operators defined as, 
\begin{eqnarray}
s_{n}^{\dag}|0\rangle&=&\frac{1}{\sqrt{2}}(|\uparrow \downarrow \rangle-
|\downarrow \uparrow \rangle)
\nonumber\\
t_{x,n}^{\dag}|0\rangle&=&\frac{-1}{\sqrt{2}}(|\uparrow \uparrow \rangle-
|\downarrow \downarrow \rangle)
\nonumber\\
t_{y,n}^{\dag}|0\rangle&=&\frac{i}{\sqrt{2}}(|\uparrow \uparrow \rangle+
|\downarrow \downarrow \rangle)
\nonumber\\
t_{z,n}^{\dag}|0\rangle&=&\frac{1}{\sqrt{2}}(|\uparrow \downarrow \rangle+
|\downarrow \uparrow \rangle)
\nonumber\\
a_{u,n,\sigma}^{\dag}|0\rangle&=&|
\sigma ,\circ \rangle_{n},\nonumber\\
a_{d,n,\sigma}^{\dag}|0\rangle&=&|
\circ ,\sigma \rangle_{n},
\end{eqnarray}
for a vertical  bond and 
\begin{eqnarray}
{s'}_{n}^{\dag}|0\rangle&=&\frac{1}{\sqrt{2}}(|\uparrow \downarrow \rangle-
|\downarrow \uparrow \rangle)
\nonumber\\
p_{x,n}^{\dag}|0\rangle&=&\frac{-1}{\sqrt{2}}(|\uparrow \uparrow \rangle-
|\downarrow \downarrow \rangle)
\nonumber\\
p_{y,n}^{\dag}|0\rangle&=&\frac{i}{\sqrt{2}}(|\uparrow \uparrow \rangle+
|\downarrow \downarrow \rangle)
\nonumber\\
p_{z,n}^{\dag}|0\rangle&=&\frac{1}{\sqrt{2}}(|\uparrow \downarrow \rangle+
|\downarrow \uparrow \rangle),
\nonumber\\
a_{l,n,\sigma}^{\dag}|0\rangle&=&|
\sigma,\circ \rangle_{n},\nonumber\\
a_{r,n,\sigma}^{\dag}|0\rangle&=&|
\circ \sigma\rangle_{n},
\label{eq:bond}
\end{eqnarray}
for a horizontal  bond at the $n$-th cluster in Fig. \ref{fig:model}.
The operators $s(s'), t_{\alpha}(p_{\alpha}) (\alpha=x, y, z)$ obey 
the bosonic commutation 
relation, while the operator $a$ the fermionic anticommutation relation. 
The notations $u,d,l,r$ specify the $up$, $down$, $left$ and $right$ sites in 
the unit cell as shown in the lower panel of  Fig. \ref{fig:model}(a).
The ket states are labeled by the 
configuration of electrons sitting at up (left) and down (right) sites
of the vertical (horizontal) bond
[$\circ$ represents an unoccupied (hole) site]. 
To restrict the Hilbert space to the 
physically sensible one, we introduce the following constraints,
\begin{eqnarray}
s_{n}^{\dag}s_{n}+\sum_{\alpha}t_{\alpha,n}^{\dag}t_{\alpha,n}+
\sum_{\sigma,i=u,d}a_{i,n,\sigma}^{\dag}a_{i,n,\sigma}=1,\\
{s'}_{n}^{\dag}{s'}_{n}+\sum_{\alpha}p_{\alpha,n}^{\dag}p_{\alpha,n}+
\sum_{\sigma,i=l,r}a_{i,n,\sigma}^{\dag}a_{i,n,\sigma}=1.
\label{eq:limit}
\end{eqnarray}    
We are now considering the dimer phase for the host spin system,
which is realized by the condensation of the singlet-bond operators,
$s_{n}$ and $s'_{n}$.  The remaining triplet
operators describe spin excitations in  the 
host system, giving  $H_{J}$ in the Fourier space,
\begin{eqnarray}
H_{J}&=&-\frac{3}{2}J_{1}N
\nonumber\\
&&+\sum_{{\bf k},\alpha}\{ 
\omega_{1k} t_{\alpha,{\bf k}}^{\dag}t_{\alpha,{\bf k}}+
\omega_{2k} p_{\alpha,{\bf k}}^{\dag}
p_{\alpha,{\bf k}} \}
\label{eq:hgp}
\end{eqnarray}
with 
\begin{eqnarray}
t_{\alpha,{\bf i}}^{\dag}=\sqrt{\frac{1}{N}}\sum_{{\bf k}}
t_{\alpha,{\bf k}}^{\dag}e^{i{\bf k}{\bf r_{\bf i}}},
\nonumber\\
p_{\alpha,{\bf i}}^{\dag}=\sqrt{\frac{1}{N}}\sum_{{\bf k}}
p_{\alpha,{\bf k}}^{\dag}e^{i{\bf k}{\bf r_{\bf i}}},
\label{eq:f-1}
\end{eqnarray}
where $\omega_{1k}=\omega_{2k}=J_1$, $N$ is the number of unit cells and 
${\bf k}$ is the wave vector. 
When we have derived the Hamiltonian (\ref{eq:hgp}), 
the interactions among triplets 
have been discarded due to their minor importance.
We note that the triplet excitations $\omega_1$ and $\omega_2$ obtained 
do not depend on the wave number,
implying  that the orthogonal-dimer structure forbids the motion of 
the triplet. In fact, it is known that the triplet excitation does not
have the dispersion up to the fifth order in the inter-dimer coupling
for the 2D model.
\cite{Miyahara}  To make our discussions more quantitative,
we use the triplet excitation energy obtained by the perturbation expansion
up to the fourth order,
\begin{eqnarray}
\omega_{1k}=J_{1}\{1-(\frac{J_{2}}{J_{1}})^{2}-\frac{1}{2}
(\frac{J_{2}}{J_{1}})^{3}
\nonumber\\
+\frac{3}{8}(\frac{J_{2}}{J_{1}})^{4}
-\frac{1}{2}(\frac{J_{3}}{J_{1}})^{2}(
\frac{J_{2}}{J_{1}})^{2}\}
\label{eq:omega1}
\\
\omega_{2k}=J_{1}\{1-(\frac{J_{3}}{J_{1}})^{2}-\frac{1}{2}
(\frac{J_{3}}{J_{1}})^{3}
\nonumber\\
+\frac{3}{8}(\frac{J_{3}}{J_{1}})^{4}
-\frac{1}{2}(\frac{J_{2}}{J_{1}})^{2}(
\frac{J_{3}}{J_{1}})^{2}\}.
\label{eq:omega2}
\end{eqnarray}

We now consider the hole dynamics in the orthogonal-dimer system. 
To this end, 
we define  the retarded Green function for
 physical fermions in the $4 \times 4$  matrix form,
\begin{eqnarray}
G_{\sigma}({\bf k},t)&=&-i\Theta(t) \langle 
D|\{\phi_{{\bf k},\sigma}(t),
\phi_{{\bf k},\sigma}^{\dag}\}|D \rangle,
\label{eq:green-2}
\end{eqnarray}    
where $\phi_{{\bf k},\sigma}^{\dag}=(\hat{c}_{u,{\bf k},\sigma}^{\dag}
,\hat{c}_{d,{\bf k},\sigma}^{\dag},\hat{c}_{r,{\bf k},\sigma}^{\dag},
\hat{c}_{l,{\bf k},\sigma}^{\dag})$ and $|D \rangle $ 
is the dimer singlet ground state.\cite{Miyahara}
We employ here the SCBA
to obtain the self-energy of the  Green functions. 
It is known that
this approximation works rather well as far as 
a single hole doped in 
quantum spin systems is concerned.\cite{rink,horsh,liu,frank,jurecka,Saito} 
Following standard procedures in the 
SCBA, we compute the spectral function of
the physical fermion defined by
\begin{eqnarray}    
A({\bf k},\omega)=-\frac{1}{\pi}  {\rm Im} \,\,
{\rm Tr}\,\, G({\bf k},\omega),
\end{eqnarray}    
which directly gives the photoemission spectrum.
This function satisfies the sum rule,
$\int A({\bf k},\omega)d\omega=2$.

This completes the formulation of the bond-operator treatment
of  the dimer phase.  We will also employ the ED calculation of the 
finite-size system to check the validity of the bond-operator
approach. Moreover, we will see that 
the latter ED calculation is  efficient 
 to discuss the  hole dynamics in another singlet phase
of the model, i.e. the plaquette phase.

Here we make a comment on the exchange couplings
of our generalized model.  If 
we consider  the {\it t-J} model (\ref{eq:hamiltonian}) is 
 to be derived  from a strong-coupling limit of the Hubbard model 
 with on-site repulsion $U$, the ratio of the parameters is given 
as,
\begin{eqnarray}
\frac{t_{1}^{2}}{J_{1}}=\frac{t_{2}^{2}}{J_{2}}
=\frac{t_{3}^{2}}{J_{3}}=\frac{t_{4}^{2}}{J_{4}}=\frac{U}{4}.
\label{eq:parameter}
\end{eqnarray}
We use these relations when  the ratio of the exchange
coupling and hopping is altered.

\section{Hole dynamics in orthogonal-dimer system}\label{sec4}

\subsection{orthogonal-dimer chain}

\begin{figure}[htb]
\begin{center}
\includegraphics[width=7cm]{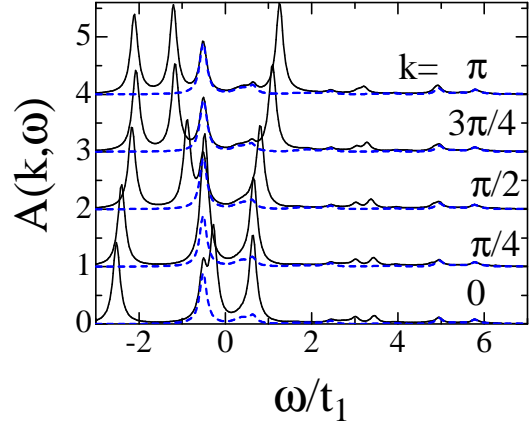}
\end{center}       
\caption{Spectral function $A({\bf k},\omega)$ for the 
orthogonal-dimer chain with $J_{1}=5$, $J_{2}=5 \times 0.635$ and
$J_3=J_4=0$ (dimer phase). The  contribution of the dispersionless mode
is shown by the dashed line. 
}
\label{fig:1d-1}
\end{figure}

We begin with 
 the 1D spin chain ($J_{3}=J_{4}=0$) shown in the lower 
panel of Fig. \ref{fig:model}(a),
which may have essential properties inherent in the
orthogonal-dimer systems.\cite{Ivanov,Schulenburg,Richter,Koga-3,Honecker} 
Before presenting the computed results, we note that 
the ground state of the chain model belongs to the dimer 
(plaquette) singlet phase for $J_{2}/J_{1}<J_c (>J_c)$
with $J_c=0.819$.\cite{Ivanov,Schulenburg,Richter,Koga-3}
 Shown in Fig. \ref{fig:1d-1} is 
 the spectral function $A({\bf k},\omega)$
for the dimer phase, which has been 
calculated by combining the bond operator 
method with the SCBA.
Since there are four sites in the unit cell,  we have 
four kinds of clear  peaks in the spectral function:
three dispersive modes and one dispersionless
mode. We refer to the lowest mode having clear dispersion
 as the quasi-particle state. We thus say that the
quasi-particle state is well defined and
 is mobile rather freely in the background of
the dimer singlets.

\begin{figure}[htb]
\begin{center}
\includegraphics[width=7cm]{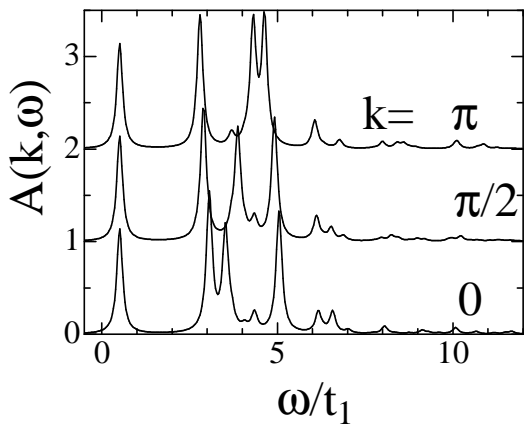}
\end{center}       
\caption{Spectral function $A({\bf k},\omega)$ for the 
chain in the plaquette phase $(J_{1}=5$, $J_{2}=5 \times 0.85)$
obtained by the ED for the small system (16 sites).
}
\label{fig:0d-3}
\end{figure}

In the plaquette phase, the spectral function $A({\bf k},\omega)$ shows
 a similar four-peak structure, as seen  in Fig. \ref{fig:0d-3},
which is obtained by the ED calculation 
 of the finite system
 $(J_{1}=5$, $J_{2}=5 \times 0.85)$.
  In this case, however, 
 the lowest quasi-particle peak is dispersionless,
in contrast to the dimer phase, implying that a 
doped hole is forbidden to itinerate in the chain system. This
results in a  completely localized quasi-particle state.

\begin{figure}[htb]
\begin{center}
\includegraphics[width=7cm]{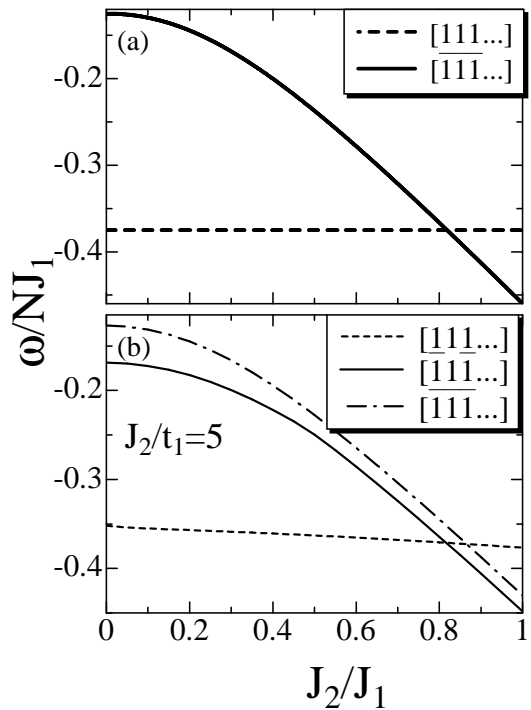}
\end{center}       
\caption{
 The energy of the lowest state in the distinct subspaces
(a) undoped case, (b) one-hole case.
The data are obtained by the ED for the system (24 sites) 
with periodic boundary conditions.
}
\label{fig:0d-2}
\end{figure}

We now clarify why the above characteristic features appear in the 
spectral function, by taking into account symmetry property
 inherent in the
orthogonal-dimer structure. This consideration may be also helpful to
discuss the 2D and 3D cases.
Notice first that the system has symmetry of space
inversion $P$ for each vertical dimer-pair; namely if the pair forms
a singlet (triplet), we have the eigenvalue $P=1$ ($P=-1$).
Then, the Hilbert space of the Hamiltonian is classified into 
each subspace specified by a set of the eigenvalues $[\{P_i\}]$.
In the dimer (plaquette) phase  for the undoped case, 
 the singlet (triplet) state is realized on 
every vertical bond.\cite{Koga-3}
Therefore the ground state of these phases is 
respectively given by 
the lowest state in the 
subspaces $[111\cdots]$ and $[\bar{1}\bar{1}\bar{1}\cdots]$,
as shown in Fig. \ref{fig:0d-2}(a). 

Now consider the hole doping effect.
In the dimer phase, 
a certain dimer singlet may be  broken by a doped hole, but
 the bonding state (symmetric superposition) is formed 
between the doped hole and the corresponding unpaired spin.
Therefore the ground state in the doped case is still in  the same 
subspace $[111\cdots]$, since the bonding state on a vertical bond has 
the eigenvalue $P=1$ ( Fig. \ref{fig:0d-2}(b)).  
This symmetry property renders the 
hole state mobile, making a well defined quasi-particle state.
Two other dispersive modes in Fig. \ref{fig:1d-1} belong to
this class of $P=1$.  On the other hand, 
a dispersionless  hole state is 
realized by  creating an anti-bonding state ($P=-1$) at 
a broken-dimer site, which thus belongs to the 
subspace $[1\bar{1}11\cdots]$.  Since the different 
inversion symmetry between neighboring pairs prohibits the
motion of a hole, leading to a completely-localized excited state.

Similar consideration can be applied to the hole doping 
in the plaquette phase.  In this case, however,
the energy level of the  anti-bonding hole state $(P=-1)$
is higher than that for the bonding state ($P=1$).
Therefore, the ground state is not realized 
in the subspace of $[\bar{1}\bar{1}\bar{1}\bar{1}\cdots]$ but in 
$[\bar{1}1\bar{1}\bar{1}\cdots]$, as indeed seen
 in Fig. \ref{fig:0d-2}(b).  Therefore, we end up with 
a completely-localized  quasi-particle (lowest-energy) state 
upon hole doping in the plaquette phase.

\subsection{2D Shastry-Sutherland model}\label{sec5}
\begin{figure}[htb]
\begin{center}
\includegraphics[width=7cm]{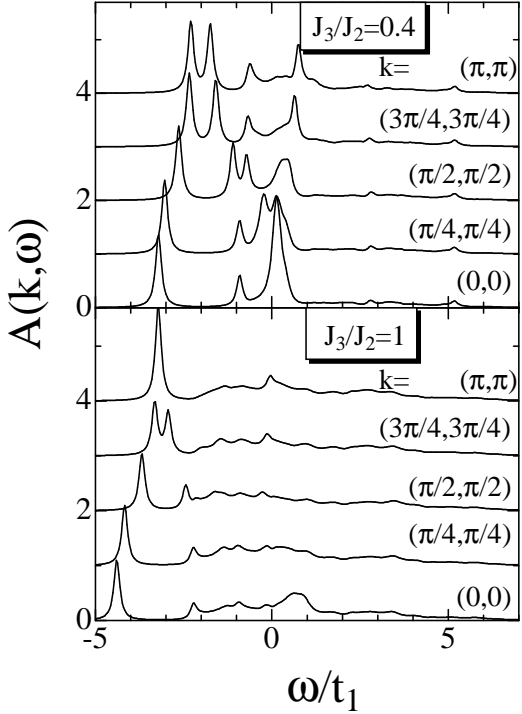}
\end{center}       
\caption{Spectral function $A({\bf k},\omega)$ 
in the dimer phase ($J_{1}=5$ and $J_{2}=J_{1}\times 0.635$).
}
\label{fig:1d-2}
\end{figure}

We now turn to the 2D case.
In the following, 
we discuss how the spectral function of a doped hole is changed 
when the interchain coupling $J_{3}$ is introduced.
We begin with  a  quasi-particle state in the dimer phase. 
The spectral function obtained by the SCBA is shown 
in Fig. \ref{fig:1d-2}. It is seen that 
the introduction of the interchain coupling 
increases the band width of the lowest mode
while  keeping its dispersive peak. This
implies  that the  quasi-particle state is well defined 
even in the 2D Shastry-Sutherland model,
in accordance  with the results of Vojta.\cite{Vojta}
On the other hand,  other peaks in the high energy region 
are considerably smeared with increase of the interchain coupling.
Such smearing effects are induced by scattering by
triplet bosons.
 
\begin{figure}[htb]
\begin{center}
\includegraphics[width=7cm]{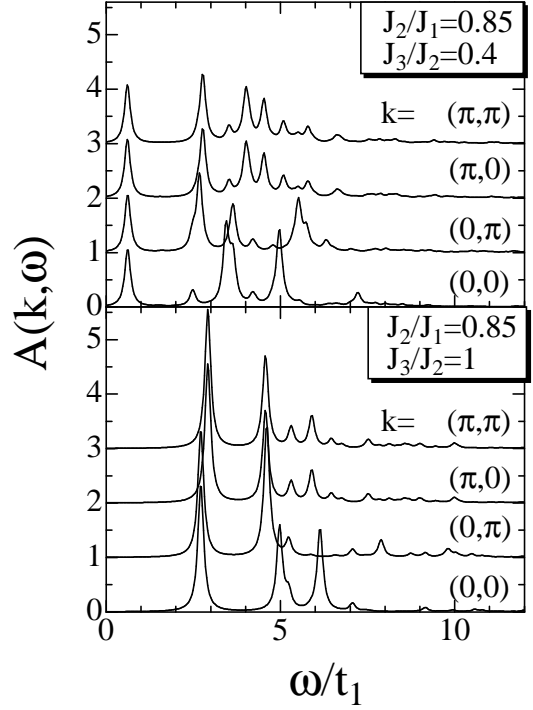}
\end{center}       
\caption{Spectral function $A({\bf k},\omega)$ 
in the plaquette phase. The choice of the parameters
are $J_{1}=5$ and $J_{2}=J_{3}=J_{1}\times 0.85$, for which 
the plaquette singlet phase should be stabilized.
\cite{Koga-1,Takushima,Sigrist2}}
\label{fig:02d-3}
\end{figure}

Now let us turn to the plaquette phase.
Note that this type of plaquette-singlet phase may exist even
in the 2D Shastry-Sutherland model 
due to strong frustration, which is believed to be 
sandwiched by the  dimer phase and the antiferromagnetically 
ordered phase.\cite{Koga-1,Takushima,Sigrist2}
We discuss here the dynamical properties of the doped hole in the
plaquette phase. 
By using the ED calculation of the $4 \times 4$ small cluster, 
we obtain
the spectral function as shown in Fig. \ref{fig:02d-3}.
The introduction of the interchain coupling $J_{3}$
has little effect on the quasi-particle state (lowest-energy state)
except for the simple energy shift.
In particular, it should be noticed that the 
dispersionless quasi-particle state can persist even in the
2D system. We also note that higher
 dispersive peaks are not so much smeared in comparison with those in the 
dimer phase. This is partially due to the finite-size
calculation we have used for the plaquette phase.

\begin{figure}[htb]
\begin{center}
\includegraphics[width=7cm]{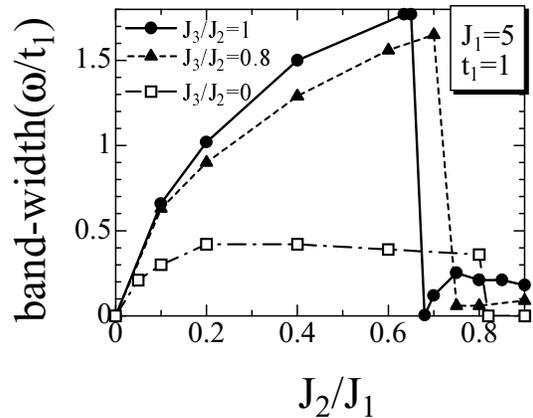}
\end{center}       
\caption{Band-width of the lowest energy band (quasi-particle
state) as a  function of $J_{2}/J_{1}$ with  $J_{3}/J_{2}$ being fixed.
}
\label{fig:02d-1}
\end{figure}

For reference,
we plot  the  band width of the quasi-particle state in
Fig. \ref{fig:02d-1}.
In the dimer phase, the band width is roughly
proportional to hopping $t$, and thus it increases in proportion 
to the square root 
of $J_{2}/J_{1}$  since the ratio of hopping and exchange 
coupling is fixed as in eq. (14). 
At the critical  point $J_{c}$, the 
band width shows discontinuity, reflecting  the first-order 
quantum phase transition.
Beyond the critical point $J_c$, 
the doped hole  hardly itinerates, as mentioned above. In particular, 
in the chain case ($J_3=0$), it is
completely localized.
By increasing the interchain coupling $J_{3}$,  
 the doped hole can itinerate in the system 
even for the plaquette phase, giving rise to  a slight increase of
 the band width.

Summarizing the above results, we can say that the one-particle 
(photo-emission) spectrum provides an efficient probe to
  distinguish the two
spin-gap phases clearly; i.e.
a dispersive (almost localized) quasi-particle 
state characterizes the dimer (plaquette)
phase.  We wish to note that  similar 
behavior appears in a triplet spin excitation spectrum in 
the undoped case, for which characteristic properties in the 
two phases are interchanged.
Namely, a triplet excitation forms a completely
localized (dispersive) mode for the dimer (plaquette) phase in the
1D case.
Even in 2D, it has been shown that a triplet excitation in the dimer
phase is almost localized.\cite{Miyahara}
Therefore, we can efficiently use these properties 
to analyze the photoemission spectrum
and/or the neutron scattering  experiments to figure out whether a plaquette 
phase can be really realized experimentally by changing the 
pressure, the magnetic field, etc.

\subsection{Effects of the interlayer coupling}
Finally, we discuss to what extent the above characteristic properties 
can persist in the presence of the inter-layer coupling.
Some recent reports claimed that the compound $\rm SrCu_2(BO_3)_2$ 
has rather large interlayer couplings, 
which may not be negligible to explain 
the experimental findings. \cite{Miyahara2,Knetter}
Nevertheless, it is known that some of physical quantities can 
be explained rather well by the 2D model.
We address this question on the hole dynamics  
 in  the quasi-2D orthogonal-dimer system.
We are mainly concerned with the dimer phase below in this section.

\begin{figure}[htb]
\begin{center}
\includegraphics[width=7cm]{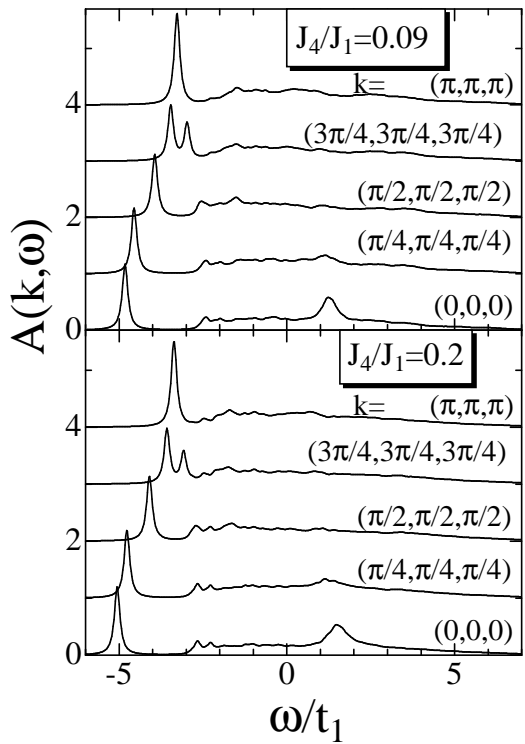}
\end{center}       
\caption{Spectral function $A({\bf k},\omega)$ 
in the dimer phase ($J_{1}=5$ and $J_{2}=J_{3}=J_{1}\times0.635$)
with the interlayer coupling $J_{4}=J_{1}\times 0.09$ and 
$J_{4}=J_{1}\times 0.2$. Both choices belong to the dimer phase.
}
\label{fig:1d-2-2}
\end{figure}

We show the computed spectral function of a doped hole in the
quasi-2D case in 
 Fig. \ref{fig:1d-2-2}, where we have chosen two different interlayer
couplings $J_{4}/J_{1}=0.09$ and $0.2$ with fixed 
$J_{2}/J_{1}=J_{3}/J_{1}=0.635$.
{\it A remarkable point 
is that the introduction of the interlayer couplings $J_{4}$ and $t_{4}$
does not obscure the spectral function}, and seems to  even sharpen its 
characteristic peak structure in some energy region.  We show
below that this remarkable behavior is caused by  
 the specific arrangement of dimers in the direction along the $c$-axis.

\begin{figure}[htb]
\begin{center}
\includegraphics[width=7cm]{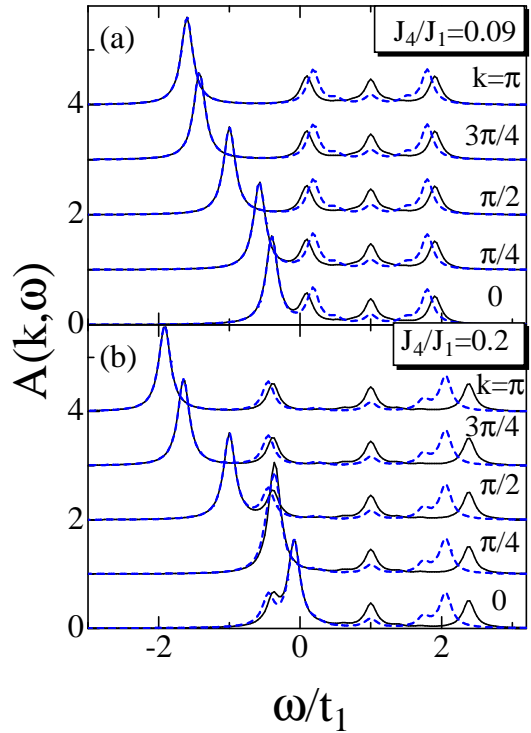}
\end{center}       
\caption{Spectral function of tetrahedra chain. (a)$J_{1}=2$ and 
$J_{4}=J_{1}\times 0.09$ (b)$J_{1}=2$ and $J_{4}=J_{1} \times 
0.2$. The dashed line represents the results obtained by 
the ED.
}
\label{fig:3d}
\end{figure}

To make this point clear, we  exploit 
a simplified model of the two-leg ladder with diagonal bonds, 
which is the basic structure along the $c$-direction 
 (see Fig. \ref{fig:model} (b)). 
Magnetic properties of 
this ladder model  have been already studied in detail.
\cite{Gelfand,Honecker,Brenig,Kim,Totsuka-lad,Sutherland,Kawaguchi} 
To  analyze the hole dynamics in this ladder system, it is 
convenient to introduce the fermionic operators for bonding and 
antibonding hole states, which are defined as
\begin{eqnarray}
a_{s,\sigma,n}^{\dag} &=&\frac{1}{\sqrt{2}}(a_{2,\sigma,n}^{\dag}+
a_{1,\sigma,n}^{\dag}),
\nonumber\\
a_{a,\sigma,n}^{\dag}&=&\frac{1}{\sqrt{2}}(a_{2,\sigma,n}^{\dag}-
a_{1,\sigma,n}^{\dag}).
\label{eq:bond1}
\end{eqnarray}
Here the fermionic operators $a_{1,\sigma,n}$ and  $a_{2,\sigma,n}$ 
denote  hole states created at two distinct sites
of the $n$-th rung in the $c$-direction.  These two states 
span two independent Hilbert spaces when the
hopping and the exchange couplings are introduced along the $c$-direction.
The  Hamiltonian for the bonding hole state 
is immediately diagonalized, and 
the corresponding  eigenstate and the eigenvalue are given as,
\begin{eqnarray}
|\psi_{s,\sigma,k} \rangle
&=&\frac{1}{\sqrt{N'}}\sum_{n}\exp(ikn)a_{s,\sigma,n}^{\dag}s_{n}|D\rangle,\\
\epsilon_{k}&=&t_{1}-2t_{4}\cos{k},
\label{eq:e.s.}
\end{eqnarray}
where $N'$ is the number of unit cells.
An important point is that the doped hole in a bonding (symmetric) state 
can move in the ladder {\it without producing any triplet excitations}. 
This is a strong constraint coming from local symmetry
due to the unique dimer alignment in the $c$-direction.
Therefore, the quasi-particle peak appears clearly in Fig. \ref{fig:3d}.
On the other hand, an anti-bonding (anti-symmetric) hole state
$a_{a}$ is completely localized, as shown in Fig. \ref{fig:3d}. 
We have also confirmed this conclusion by calculating
the spectral function by the ED for
the small cluster $N'=8$, which is shown in Fig. \ref{fig:3d}
by the dashed line. It is seen that the  results show
good agreement with those obtained by the SCBA.

The above  analysis is valid  for the ladder system, but
most of essential features
 may appear even in the quasi-2D case along the 
$c$-direction although
the profile of the spectrum  may be somewhat obscured.
This naturally explains  why the quasi-particle peak 
in the low energy region 
in the orthogonal-dimer system is not smeared by the interlayer coupling. 


To conclude this section, we check how well
the quasi-particle state is stabilized in 3D
by estimating the quasi-particle weight defined by
\begin{eqnarray}
Z({\bf k})=\left (1-Re \frac{\partial \Sigma({\bf k}, \omega)}
{\partial \omega} \right )^{-1}|_{\omega=\epsilon_{{\bf k}}},
\label{eq:reno}
\end{eqnarray}
where $\epsilon_{{\bf k}}$ is the energy of the quasi-particle. 
\begin{figure}[htb]
\begin{center}
\includegraphics[width=7cm]{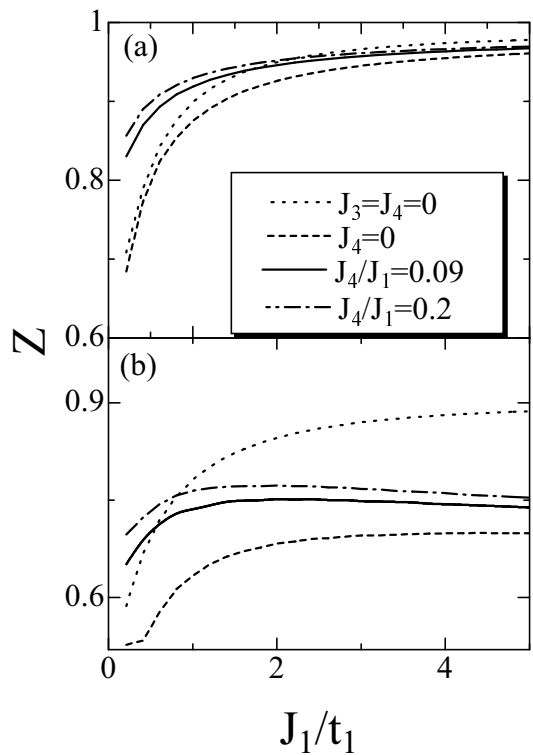}
\end{center}       
\caption{Renormalization factor at the $\Gamma$ point with 
(a)$J_{2}/J_{1}=J_{3}/J_{1}=0.2$ and 
(b)$J_{2}/J_{1}=J_{3}/J_{1}=0.635$. 
}
\label{fig:1d-3}
\end{figure}
Shown in  Fig. \ref{fig:1d-3} is 
 the  weight computed at the $\Gamma$ point
when the ratio of the exchange coupling and the hopping is varied.
In the parameter region shown in the figure, there are
two characteristic features. (i) the change in character from
1D to 2D: By comparing the data of $J_3=J_4=0$ (dotted line)
with those of $J_4=0$ (broken line), we notice that the increase of the 
interchain coupling results in the decrease of the
quasi-particle weight in accordance with the discussions
given above. (ii) Change from 2D to 3D: 
On the other hand, the increase of the inter-layer 
coupling does not lead to the 
 decrease of the weight, but  has a tendency to
increase it, as seen from the solid and dash-dotted lines 
in Fig. \ref{fig:1d-3}. This property is related to the fact that 
the motion of a hole in the $c$-direction is quite free, rendering 
the quasi-particle state more stable.


\section{Summary}\label{Summary}

We have investigated the hole dynamics in the 
orthogonal-dimer model proposed 
for the quasi-2D compound $\rm SrCu_{2}(BO_{3})_{2}$.
In particular, the characteristic properties present
in the one-particle spectrum have been 
clarified in detail, by 
systematically studying the one, two and three dimensional
 systems.  We have demonstrated that
the quasi-particle state of a doped hole is 
well stabilized due to the specific orthogonal-dimer
structure. Furthermore, it has been found that 
the one-particle spectral function  provides an efficient 
way to distinguish the two spin-gap phases; i.e.
a dispersive (almost localized) quasi-particle mode characterizes 
the dimer (plaquette) phase.  This remarkable property
is inherent in the orthogonal-dimer system, which stems 
from a strong constraint due to local symmetry.

We expect that the plaquette singlet 
phase could be realized experimentally by applying the pressure 
or the chemical pressure with substitution of certain  elements.
In order to judge which type of spin-gap phase is realized in such case,
 an efficient probe is necessary to distinguish them
properly. We have found here that the photoemission spectrum
can be an efficient method to specify the nature of the spin-gap phases.

\section{Acknowledgments}
Our ED computational programs are based on TITPACK, Version 2, 
by Nishimori. 
This work was partly supported by a Grant-in-Aid from the Ministry 
of Education, Science, Sports and Culture of Japan. 
A part of calculations was done at the Supercomputer Center 
at the Institute for Solid State Physics, University of Tokyo.

%



\begin{thebibliography}{99}

\bibitem{Kageyama}
H. Kageyama, K. Yosimura, R. Stern, N. V. Mushnikov, K. Onizuka, 
M. Kato, K. Kosuge, C. P. Slichter, T. Goto 
and Y. Ueda, 
Phys. Rev. Lett. {\bf 82}, 3168 (1999).

\bibitem{Kodama}
K. Kodama, M. Takigawa, M. Horovatic, C. Berthier, H. Kageyama, 
Y, Ueda, S. Miyahara, F. Becca and F. Mila, 
Science {\bf 298} 395 (2002)

\bibitem{Onizuka}
K. Onizuka, H. Kageyama, Y. Narumi, K. Kindo, 
Y. Ueda and T. Goto, 
J. Phys. Soc. Jpn. {\bf 69}, 1016 (2000).

\bibitem{Aso}
N. Aso, H. Kageyama, K. Nukui, M. Nishi, K. Kakurai, K. Onizuka,
Y. Ueda, and H. Kadowaki, to be published in J. Phys. Chem. Solid. (2002)

\bibitem{Nojiri}
H. Nojiri, H. Kageyama, K. Onizuka, Y. Ueda and 
M. Motokawa,  
J. Phys. Soc. Jpn. {\bf 68}, 2906 (1999).

\bibitem{Lemmens}
P. Lemmens, M. Grove, M. Fischer, G. G\"untherodt, V. N. Kotov,
H. Kageyama, K. Onizuka and Y. Ueda, Phys. Rev. Lett. {\bf 85}, 2605 (2000).

\bibitem{Kageyama2}
H. Kageyama, N. V. Mushnikov, M. Yamada, 
T. Goto and Y. Ueda, 
Physica B {\bf 329}, 1020 (2003).

\bibitem{Miyahara}
S. Miyahara and K. Ueda, 
Phys. Rev. Lett. {\bf 82}, 3701 (1999).  

\bibitem{Shastry}
B. S. Shastry and B. Sutherland, Physica B {\bf 108}, 1069 
(1981).

\bibitem{Miyahara-4}
S. Miyahara and K. Ueda, 
J. Phys: Condens. Matter {\bf 15}, R327-R366 (2003)

\bibitem{Koga-1}
A. Koga and N. Kawakami, 
Phys. Rev. Lett. {\bf 84}, 4461 (2000).  

\bibitem{Knetter}
C. Knetter, A. $\rm B\ddot{u} hler$, E. $\rm M\ddot{u} ller-Hartmann$ 
and G. S. Uhrig, 
Phys. Rev. Lett. {\bf 85}, 3958 (2000)

\bibitem{Weihong}
Z. Weihong, C. J. Harmer and J. Oitmaa, 
Phys. Rev. B {\bf 60}, 6608 (1999);
Phys. Rev. B {\bf 65}, 014408 (2002).

\bibitem{Muller}
E. $\rm M\ddot{u} ller-Hartmann$, R. R. P. Singh, C. Knetter and 
G. Uhrig, 
Phys. Rev. Lett. {\bf 84}, 1808 (2000).

\bibitem{Mila}
M. Albrecht and F. Mila, 
Europhys. Lett. {\bf 34}, 145 (1996)

\bibitem{Chung}
C. H. Chung, J. B. Marston and S. Sachdev, 
Phys. Rev. B {\bf 64}, 134407 (2001).

\bibitem{Carpentier}
D. Carpentier and L. Balents, 
Phys. Rev. B {\bf 65}, 024427 (2002).

\bibitem{Takushima}
Y. Takushima, A. Koga and N. Kawakami, 
J. Phys. Soc. Jpn. {\bf 70}, 1369 (2000).

\bibitem{Sigrist2}
A. $\rm L\ddot{a}u chli$, S. Wessel and M. Sigrist
Phys. Rev. B {\bf 66}, 014401 (2002).

\bibitem{TM}
K. Totsuka, S. Miyahara and K. Ueda,
Phys. Rev. Lett. {\bf 86}, 520 (2001).

\bibitem{Momoi}
T. Momoi and K. Totsuka, 
Phys. Rev. B {\bf 61}, 3231 (2000);
Phys. Rev. B {\bf 62}, 15067 (2000).

\bibitem{Totsuka}
K. Totsuka, S. Miyahara and K. Ueda, 
Phys. Rev. Lett. {\bf 86}, 520 (2000)

\bibitem{Fukumoto-2}
Y. Fukumoto, J. Phys. Soc. Jpn.{\bf 69}, 2755 (2000).

\bibitem{Knetter-2}
C. Knetter and G. Uhrig, 
cond-mat/0309408.

\bibitem{Munehisa}
T. Munehisa and Y. Munehisa, 
J. Phys. Soc. Jpn. {\bf 69}, 1286 (2000).

\bibitem{Miyahara3}
S. Miyahara and K. Ueda, 
Phys. Rev. B {\bf 61}, 3417 (2000).

\bibitem{Misguich}
G. Misguich, Th. Jolicoeur and S. M. Girvin, 
Phys. Rev. Lett. {\bf 27}, 097203 (2001)

\bibitem{Fukumoto-pla}
Y. Fukumoto and A. Oguchi, J. Phys. Soc. Jpn. {\bf 69}, 1286 (2000).

\bibitem{Miyahara7}
S. Miyahara, F. Becca and F. Mila, 
Phys. Rev. B {\bf 68}, 024401 (2003).

\bibitem{Miyahara2}
S. Miyahara and K. Ueda, 
J. Phys. Soc. Jpn. Suppl. B {\bf 69}, 72 (2000).

\bibitem{Vojta}
M. Vojta, Phys. Rev. B {\bf 61}, 11309 (2000).

\bibitem{Dope} 
H. Kageyama, Y. Narumi, K. Kindo, K. Onizuka, Y. Ueda, and T. Goto, 
J. Alloys and Compounds {\bf 317-318}, 177-182 (2001).  

\bibitem{rink}
S. Schmitt-Rink , C. M. Varma and A. E. Ruckenstein, 
Phys. Rev. Lett. {\bf 60}, 2793 (1988).

\bibitem{horsh}
G. Martinez and P. Horsh, Phys. Rev. B {\bf 44}, 317 (1991).

\bibitem{liu}
Z. Liu and E. Manousakis, Phys. Rev. B {\bf 45}, 2425 (1992).

\bibitem{frank}
F. Magsiglio, A. Ruckenstein, Schumitt-Rink and C. Varma, 
Phys. Rev. B {\bf 43}, 10882 (1991).

\bibitem{jurecka}
C. Jurecka and W. Brenig, 
Phys. Rev. B {\bf 63}, 094409 (2001).

\bibitem{Saito}
Y. Saito, A. Koga and N. Kawakami
J. Phys. Soc. Jpn. {\bf 72}, 60349 (2003).

\bibitem{Sachdev}
S. Sachdev and N. Bhatt, 
Phys. Rev. B {\bf 41}, 9323 (1990).

\bibitem{Sigrist}
S. Gopalan, T. M. Rice and M. Sigrist, 
Phys. Rev. B {\bf 49}, 8901 (1994).

\bibitem{Matsumoto}
M. Matsumoto, B. Normand, T. M. Rice and M. Sigrist, 
Phys. Rev. Lett. {\bf 89}, 077203 (2002)

\bibitem{Ueda}
K. Ueda and S. Miyahara, 
J. Phys. Condens. Matter. {\bf 11} L175 (1999).

\bibitem{Koga-2}
A. Koga, 
J. Phys. Soc. Jpn. {\bf 69}, 3509 (2000).

\bibitem{Ivanov}
N. B. Ivanov and J. Richter,
Phys. Lett. A {\bf 232}, 308 (1997).

\bibitem{Schulenburg}
J. Schulenburg and J. Richter, 
Phys. Rev. B {\bf 65}, 054420 (2002);
Phys. Rev. B {\bf 66}, 134419 (2002).

\bibitem{Richter}
J. Richter, N. B. Ivanov and J. Schulenburg, 
J. Phys. Condens. Matter. {\bf 10} 3635 (1998).

\bibitem{Koga-3}
A. Koga, K. Okunishi and N. Kawakami, 
Phys. Rev. B {\bf 62}, 5558 (2000);
Phys. Rev. B {\bf 65}, 214415 (2002).

\bibitem{Honecker}
A. Honecker, J. Schulenburg and J. Richter, 
cond-mat/0309425.

\bibitem{Gelfand}
M. P. Gelfand, 
Phys. Rev. B {\bf 43}, 8644 (1991).

\bibitem{Honecker}
A. Honecker, F. Mila and M. Troyer, 
Eur. Phys. J. B. {\bf 15}, 227 (2000).

\bibitem{Brenig}
W. Brenig and K. W. Becker, 
Phys. Rev. B {\bf 64}, 214413 (2001).

\bibitem{Kim}
E. H. Kim, G. $\rm F \acute{a} th$, J. $\rm S \acute{o}lyom$ and 
D. J. Scalapino, 
Phys. Rev. B {\bf 62}, 14965 (2000).

\bibitem{Totsuka-lad}
K. Totsuka and H.-J. Mikeska, Phys. Rev. B {\bf 66}, 054435 (2002).

\bibitem{Sutherland}
B. Sutherland, Phys. Rev. B {\bf 62}, 11499 (2000).

\bibitem{Kawaguchi}
A. Kawaguchi, A. Koga, K. Okunishi and N. Kawakami
J. Phys. Soc. Jpn. {\bf 72}, 405 (2003).






\end{thebibliography}
\end{document}